\documentclass[12pt]{article}

\usepackage{amsmath}
\usepackage{epsfig}

\begin{document}

\title{Photoionization of the bound systems at high energies}
\author{E. G. Drukarev, A. I. Mikhailov\\
National Research Center "Kurchatov Institute, \\B. P. Konstantinov Petersburg Nuclear Physics Institute, Gatchina,\\St. Petersburg 188300, Russia }
\maketitle

\begin{abstract}
 We consider photoionization of a system bound by the central potential $V(r)$. We demonstrate that the high energy nonrelativistic asymptotics of the photoionization cross section can be obtained without solving the wave equation. The asymptotics can be expressed in terms of the Fourier transform of the potential by employing the Lippmann--Schwinger equation. We find the asymptotics for the screened Coulomb field. We demonstrate that the leading corrections to this asymptotics are described by the universal factor. The high energy nonrelativistic asymptotics is found to be determined by the analytic properties of the potential $V(r)$. We show that the energy dependence of the asymptotics of photoionization cross sections of fullerenes is to large extent model dependent. We demonstrate that if the fullerene field $V(r)$ is approximated by the function with singularities in the complex plane, the power drop of the asymptotics is reached at the energies which as so high that the cross section becomes unobservably small. The preasymptotic behavior with a faster drop of the cross sections becomes important in these cases.

\end{abstract}

\section{Introduction}
In this paper we analyze the ionization of a bound system by the photons carrying the energy $\omega$ which exceeds strongly the binding energy of the system $I_B>0$. We calculate the high energy nonrelativistic asymptotics of the cross sections. In other words we find the leading terms of their expansions in powers of $\omega^{-1}$ at
\begin{equation}
I_B \ll \omega \ll m.
\label{0}
\end{equation}
Here $m$ is the electron rest mass (we employ the relativistic system of units with  $\hbar=1$, $c=1$, the squared electron charge is $e^2=\alpha=1/137$). We demonstrate that the asymptotics of the cross section $\sigma(\omega)$ can be obtained without solving the wave equation. We find also the energy behavior of the cross sections before the asymptotics is reached.

The ratio of the  photoionization cross sections at two high values $\omega_1$ and $\omega_2$ for any $s$ state in the system bound by the $V(r)$,
can be expressed by the simple equation
\begin{equation}
\frac{\sigma(\omega_1)}{\sigma(\omega_2)}=\Big(\frac{\omega_2}{\omega_1}\Big)^{3/2}\frac{|V(p_1)|^2}{|V(p_2)|^2},
\label{0a}
\end{equation}
with $p_i=\sqrt{2m\omega_i}$ the moments of photoelectrons.
Thus the cross sections ratio is presented in terms of the Fourier transform
\begin{equation}
\tilde V(p)=\int d^3r V(r)e^{-i{\bf p}{\bf r}}
\label{2}
\end{equation}
of the potential $V(r)$. We omit the tilde sign for the Furier transform further.

The formula  (2) is based on the plane wave description of the photoelectrons in photoionization of $s$ states. This becomes possible if the operator of interaction between the electron and photon in the velocity form ("gauge") is employed \cite{1}.
Thus the high energy photoionization cross section can be expressed in terms of the Fourier transform of the bound state function $\psi(p)$
\begin{equation}
\sigma(\omega)=\frac{4\alpha p}{3}n_e|\psi(p)|^2.
\label{9b}
\end{equation}
Here $p$ is the photoelectron momentum,  $n_e$ is the number of electrons in ionized $s$ state.
Note that due to the energy conservation law
\begin{equation}
\omega=\frac{p^2}{2m}+I_B,
\label{2a}
\end{equation}
and due to the condition $\omega \gg I_B$ the photoelectron momentum exceeds strongly the characteristic momentum of the bound state $\mu=\sqrt{2mI_B}$
\begin{equation}
p \gg \mu.
\label{2b}
\end{equation}
This enables to present the function
$\psi(p)$ in terms of the potential $V(p)$ by using the Lippmann-Schwinger equation \cite{2} as
\begin{equation}
\psi(p)=-\frac{1}{\omega}J(p),
\label{3}
\end{equation}
with
\begin{equation}
J(p)=\int\frac{d^3f}{(2\pi)^3}V({\bf p}-{\bf f})\psi({\bf f}).
\label{3d}
\end{equation}

The integral is saturated by small $f \sim \mu \ll p$ if $pV(p) \rightarrow 0$ at $p \rightarrow \infty$. Thus the shape of the function $\psi(p)$ is determined by that of $V(p)$.
We see that for the potentials $V(p)$ which can be expanded in powers of $ 1/p$ at large $p$
\begin{equation}
J(p)=V(p)\psi(r=0).
\label{3x}
\end{equation}
In these cases we find a simple equation for the asymptotic cross sections \cite{2a}
\begin{equation}
\sigma(\omega)=n_e\frac{4\alpha}{3}\frac{p}{\omega^2}|V(p)|^2\psi^2(r=0).\quad p=\sqrt{2m\omega},
\label{1}
\end{equation}

For large number of potentials we find
\begin{equation}
J(p)=V(p)\kappa,
\label{3xy}
\end{equation}
where the factor $\kappa$ does not depend on $p$ and is determined by the parameters of the bound state. In these cases one can find not only the cross sections ratios, but also the cross sections themselves
\begin{equation}
\sigma(\omega)=\frac{4\alpha}{3}\frac{p}{\omega^2}|V(p)|^2n_e\kappa^2.
\label{1x}
\end{equation}
 Note that calculation of $\kappa$ is not always possible. Sometimes one can only give an estimation of its magnitude.
 In such cases Eq.(\ref{1x}) provides the energy dependence of the cross section and the estimation of its numerical magnitude.

 Thus the shape of the energy dependence of the high energy photoionization cross section is determined by the shape of the potential $V(p)$. The latter is determined by the analytical properties of the potential $V(r)$ \cite{3}. Hence the asymptotics of the photoionization cross section $\sigma(\omega)$ depends on the analytical properties of the function $V(r)$.

The paper has the following structure. In Sec 2 we recall how the asymptotic analysis for the photoionization of atoms is carried out. In Sec.3 we obtain Eqs.({\ref 3}) and (\ref{1}). In Sec.4 we consider the screened Coulomb field. We find the asymptotics of the photoionization cross sections and obtain also their preasymptotic behavior. In Sec.5 we find the connection between the asymptotic behavior of the cross sections and the analytical properties of the potential. In Sec.6
we analyze the high energy photoionization of fullerenes. Here the model potentials are often employed \cite{3a},\cite{3b}, \cite{4}. We summarize in Sec.7.

\section{Asymptotic analysis for atoms}

Start with the case of the Coulomb field.  The analytical expressions for the cross section are well known \cite{1}, \cite{4}. The cross section for photoionization of the ground state in the field of the nucleus with the charge $Z$ can be written as
\begin{equation}
\sigma(\omega)=\frac{16\sqrt{2}\pi}{3Z^2}\alpha r_0^2\Big(\frac{m\alpha^2Z^2}{\omega}\Big)^{7/2}\Phi(p).
\label{4}
\end{equation}
Here $r_0=1/m\alpha$ is the Bohr radius. If the photoelectron is described by the plane wave,
$\Phi(p)=1$ and Eq.(\ref{4}) provides the well known  $\omega^{-7/2}$ drop of the cross section \cite{1}.

The explicit form of the function
$$\Phi(p)=D(\pi\xi)\exp{(4\xi\arctan\xi)}; \quad D(\pi\xi)=\frac{2\pi\xi\exp{(-2\pi\xi)}}{1-\exp{(-2\pi\xi)}},$$
with $\xi=m\alpha Z/p$ the parameter of interaction between the photoelectron and the nucleus \cite{1}
enables to calculate the leading correction to the asymptotics.  The function $\Phi(p)$ depends on the parameters $\pi\xi$ and $\xi^2$ separately, i.e., $\Phi(p)=\Phi(\pi\xi,\xi^2)$. This enables us to neglect the corrections of the order
$\xi^2$ while we include the lowest order of expansion in powers of $\pi\xi$. This gives
\begin{equation}
\Phi(\pi\xi, \xi^2=0)=1-\pi\xi.
\label{5}
\end{equation}
Here the second term on the right-hand side provides the lowest correction to the asymptotics. It makes less than $10\%$ at the photon energies $\omega > 14$ keV for $Z=1$.
For $Z=2$ it diminishes the cross section by  $10\%$ at $\omega \approx 56$ keV.
However the relativistic correction $\omega/m > 0.1$ at such energies. Hence, there is no region where the cross section takes its asymptotic form if $Z>2$.

However the deviations of the cross sections from the asymptotic law can be described by the universal Stobbe factor $S(\omega)=\Phi(\pi\xi, \xi^2=0)$ in Eq.(13) \cite{5}.
The dependence of the cross sections on the parameter $\pi\xi$ is expressed by the factor
\begin{equation}
S(\omega)=\frac{\pi\xi}{sh(\pi\xi)}\exp{(-\pi\xi)} \approx \exp{(-\pi\xi)}; \quad \xi=m\alpha Z/(2m\omega)^{1/2}.
\label{6}
\end{equation}

The photoionization cross sections ratio at high energies $\omega_1$ and $\omega_2$ is given by the relation
\begin{equation}
\frac{\sigma(\omega_1)}{\sigma(\omega_2)}=\Big(\frac{\omega_2}{\omega_1}\Big)^{7/2}\frac{S(\omega_1)}{S(\omega_2)},
\label{7}
\end{equation}
for any $s$ state.

 The experimental data on high energy photoionization of the multielectron atoms (excluding the cases when the intershell correlations should be included \cite{4}), as well as the results of the Hartree-Fock calculations are well described by Eq. (16)\cite{5}. This is not surprising since the Stobbe factor is formed at small distances of the order $1/p$ from the nucleus. Thus interactions between the atomic electrons do not change it noticeably.

\section{High energy photoionization cross section}

The photoionization cross section of a bound system can be written as
\begin{equation}
d\sigma=n_e\frac{mp}{(2\pi)^2}|F|^2d\Omega.
\label{8}
\end{equation}
Here $p$  and $\Omega$-are the photoelectron momentum and the solid angle, $F$ is the  amplitude of the process. It is assumed that the averaging over the photon polarizations is carried out. Now we calculate the amplitude $F$.

 The photoionization of a free electron is impossible. Hence the source of the field should obtain large momentum  ${\bf q}={\bf k}-{\bf p}$ with  ${\bf k}$ denoting the photon momentum. Note that $|{\bf k}|=\omega$. The energy conservation law ({\ref{2a}) requires that $p\gg k$, and $|{\bf q}| \approx p$. The recoil momentum $p$ can be transferred either by the bound electron or by the photoelectron. In the former case photoelectron is described by plane wave. The corresponding contribution to the amplitude takes the form \cite{4}
\begin{equation}
F_a=N(\omega)\frac{{\bf e}\cdot {\bf p}}{m}\psi(p); \quad N(\omega)=\Big(\frac{4\pi\alpha}{2\omega}\Big)^{1/2}.
\label{9}
\end{equation}
Here ${\bf e}$ is the polarization vector of the photon.

Now we include the contribution in which the recoil momentum is transferred by the photoelectron.
This mechanism is often called "the final state interaction" (f.s.i.).
We demonstrate that the f.s.i. contribution provides the corrections of the order $1/p$ to the amplitude $F_a$.
In the lowest order of perturbation theory contribution of the f.s.i. to the amplitude is
$$F_b=N(\omega)\int\frac{d^3f}{(2\pi)^3}\langle {\bf p}|VG(\omega+\varepsilon_B)|{\bf f}\rangle\frac{{\bf e}\cdot {\bf f}}{m}\langle {\bf f}|\psi\rangle,$$
where $G$ is the electron propagator of free motion with the matrix elements
$$ \langle {\bf f}_1|G(\varepsilon)|{\bf f}_2\rangle=g(\varepsilon, f_1)\delta({\bf f}_1-{\bf f}_2); \quad g(\varepsilon, f_1)=\frac{1}{\varepsilon-f_1^2/2m}.$$
One can write
\begin{equation}
F_b=N(\omega)\int\frac{d^3f}{(2\pi)^3}\langle {\bf p}|V|{\bf f}\rangle g(\omega+\varepsilon_B, f))\frac{{\bf e}\cdot {\bf f}}{m}\langle {\bf f}|\psi\rangle.
\label{9a}
\end{equation}
On the other hand, the amplitude $F_a$ can be presented as
$$F_a=������
N(\omega)\frac
{{\bf e}\cdot {\bf p}}{m}g(\varepsilon_B, p)
\int\frac{d^3f}{(2\pi)^3}\langle {\bf p}|V|{\bf f}\rangle
\langle {\bf f}|\psi\rangle.$$
Here the integral on the right-hand side is determined by small $f \sim \mu$ if $pV(p) \rightarrow 0$ at $p \rightarrow \infty$. Thus $F_b \sim (\mu/p)F_a \ll F_a$.
One can see the higher terms in $V$ to be quenched by additional powers of  $1/p$. Thus the asymptotic of the amplitude $F=F_a+F_b$ is determined by the contribution
$F_a$. Hence we can put $F=F_a$. We obtain Eq.(\ref{9b}) carrying out the angular integration in Eq.({\ref 8}).

Note that the term $F_b$ provides a small contribution to the amplitude $F$ due to the velocity form of the operator of interaction between the electron and photon. In the length form operator ${\bf e}{\bf p}/m$ in expressions for the amplitude $F_a$ is changed to $-\omega {\bf e}{\bf \nabla_p}$. Operator ${\bf e}{\bf f}/m$ in the f.s.i. amplitude $F_b$ is changed to $-\omega {\bf e}{\bf \nabla_f}$.
In the length form there are no arguments which would allow  to neglect $F_b^{r}$ (the upper index  $r$ denotes the  length form of the $e-\gamma$ interaction). The relative magnitudes of the amplitudes $F_a^{r}$  and $F_b^{r}$ can be found only after additional analysis. For example, for the atoms  $V(p)=-4\pi\alpha Z/p^2$ at
large $p$. Direct calculation provides $F_a^{r}=2F_a$, $F_b^{r}=-F_a$. Thus the amplitudes $F_a^{r}$ � $F_b^{r}$ are of the same order of magnitude.
The higher order f.s.i. terms are quenched by the higher powers of $1/p$. Thus in the length form the asymptotics of the amplitude is
$F^r=F_a^{r}+F_b^{r}$. It includes the lowest order f.s.i. term (see Subsec. 7.2.3 in the book \cite{4}). As expected, $F^r=F$.

Now we calculate the function $\psi(p)$ for $p \gg \mu$ employing the Lippmann--Schwinger equation
\begin{equation}
\psi=\psi_0+G(\varepsilon_B)V\psi.
\label{10}
\end{equation}
in momentum space.
Here $\psi_0$ is the wave function at $V=0$, $\varepsilon_B=-I_B<0$-is the electron energy in the bound state.
For the bound states $\psi_0=0$, and we find
$$\psi(p)=\langle {\bf p}|GV|\psi\rangle=g(\varepsilon_B, p)J(p),$$
where
$J(p)$ is determined by Eq.(\ref{3d}).
One can put $g(\varepsilon_B, p)=-2m/p^2$ at large $p$. This brings us to Eq.(\ref{3}).

One can separate three regions of the values of momentum $f$ in the integral $J(p)$. In the region $f \sim \mu \ll p$, we can put $V({\bf p}-{\bf f})=V({\bf p})$.
Hence it provides a contribution of the order $V(p)/p^2$ to the function $\psi(p)$. The  large moments $f \sim p$ for which $|{\bf p}-{\bf f}| \ll p$  provide the contribution of the order $\psi(p)/p^2$ to the right hand side of Eq.(7). It is beyond the asymptotics.
As to the region of large moments $f \sim p$ ; $|{\bf p}-{\bf f}| \sim p$,
its contribution to the right-hand side of Eq.(7)) is of the order $pV(p)\psi(p)$.

Hence the integral $J(p)$ is determined by the region of small moments  $f \sim \mu \ll p$ for the potentials $V(p)$, which drop faster than $1/p$ at large $p$.
The main contribution to $J(p)$ comes from large moments $f \sim p$ for the potentials which drop as $1/p$ if they do not contain oscillating factors. Recall that the potentials $V(p)$ which drop slower than $1/p$ correspond to the potentials $V(r)$ which increase faster than $1/r^2$ at
 $r \rightarrow 0$. They lead to the "fall on the center" phenomena \cite{6} and can not form a bound state.

Thus the integral $J(p)$ is determined by small $f \ll p$ if $pV(p) \rightarrow 0$ at $p \rightarrow \infty$. If also $V(p)$ can be expanded in powers of
$1/p$ at large $p$, we can put ${\bf p}-{\bf f}={\bf p}$ in Eq. (\ref{3d}). This provides Eq.({\ref{3x}).

We shall see in Sec.5 and 6 that $J(p)$ can be obtained in some of the cases which are of practical interest , when the discussed conditions for $V(p)$ are not true. Here we just note that sometimes the coordinate form for $J(p)$
\begin{equation}
J(p)=\int{d^3r}V(r)\psi(r)e^{-i{\bf p}{\bf r}}=\frac{4\pi}{p}\int_0^{\infty}drV(r)\chi(r)\sin{(pr)},
\label{14}
\end{equation}
with $\chi(r)=r\psi(r)$ becomes useful.

\section{Screened Coulomb potentials}

Note first that Eq.({\ref{1}) describes the asymptotics of the photoionization cross section (Eq.({\ref 4}) with $\Phi(p)=1$) for  $1s$
electron in the Coulomb field. We found it without solving the wave equation. We employed only the well known value the $\psi^2(r=0)=(m\alpha Z)^3/\pi$.

One of the possibilities to describe electron in the multielectron atom is to introduce a screened potential which behaves as $1/r$ at $r \rightarrow 0$.
The asymptotics of the photoionization cross sections in such fields is determined by behavior of the Fourier transforms of these potentials at large $p$.
Start with the Yukawa potential
\begin{equation}
V(r)=-\frac{ge^{-\lambda r}}{r}; \quad g>0,
\label{15}
\end{equation}
with
\begin{equation}
V(p)=-\frac{4\pi g}{p^2+\lambda^2}.
\label{16}
\end{equation}
At large $p \gg \lambda$ we obtain
\begin{equation}
V(p) \approx \frac{-4\pi g}{p^2}(1-\lambda^2/p^2).
\label{16a}
\end {equation}

 The asymptotics of photoionization cross section is the same as in the Coulomb potential with $\alpha Z=g$. This is because the asymptotics is determined by small distances $ r \sim 1/p \ll 1/\lambda$ where one can put $e^{-\lambda r}=1$ in Eq. ({\ref{15}). We kept the second term on the right-hand side to demonstrate that the screening corrections are proportional to $\lambda^2$ and drop as $1/\omega$.

The field created by a heavy atom can be approximated by the Thomas--Fermi potential. In the Tietz parametrization \cite {T} it takes the form
\begin{equation}
V(r)=-\frac{\alpha Z}{r}\frac{1}{(1+ar/r_0)^2}; \quad a=c_TZ^{1/3},
\label{17}
\end{equation}
with $c_T \approx 0.6$. Recall that  $r_0=1/m\alpha$ is the Bohr radius. The Fourier transform of this potential is
\begin{equation}
V(p)=-4\pi \alpha ZI(p); \quad I(p)=\frac{1}{p}\int_0^{\infty}dr\frac{\sin{pr}}{( 1+ar/r_0)^2  }=b^2\int_0^{\infty}dr\frac{\cos{pr}}{r+b}.
\label{18}
\end{equation}
with $b=r_0/a$. One can present $I$ as a power expansion in parameter $bp$. The two lowest terms of the expansion provide
\begin{equation}
I(p)=\frac{1}{p^2}(1-\frac{6a^2}{p^2b^2}).
\label{19}
\end{equation}
Hence the asymptotics of $V(p)$ is the same as for the Coulomb field with the nuclear charge $Z$. The leading corrections are of the order $1/p^2$.

These results are true for any potential with the Coulomb short distance asymptotics. To demonstrate this note that since the large $p$ behavior of
the potential $V(p)$ is determined by small $r$, we can write
\begin{equation}
V(r)=\frac{-\alpha Z}{r}(1+c_1r+c_2r^2)
\label{20}
\end{equation}
Present now $V(p)=-\alpha Z[V_0(p)+c_1V_1(p)+c_2V_2(p)]$ with the three contributions corresponding to the three terms on the right-hand side of Eq. (\ref{20}).
Calculating
\begin{equation}
V_0(p)=lim|_{\lambda \rightarrow 0}\int d^3r \frac{e^{-\lambda r}}{r}e^{-i{\bf p}{\bf r}}=\frac{4\pi}{p^2+\lambda^2}|_{\lambda=0};
\label{21}
\end{equation}
$$ V_k(p)=lim|_{\lambda\rightarrow 0}\int d^3r r^k\frac{e^{-\lambda r}}{r}e^{-i{\bf p}{\bf r}}=(-\frac{\partial}{\partial \lambda})^k\frac{4\pi}{p^2+\lambda^2}|_{\lambda=0}. $$
we obtain $V_0=4\pi/p^2$, $V_1=0$, $V_2=-8\pi/p^4$. This are the Coulomb asymptotics and the leading corrections of the order $1/p^2$.

Let us clarify the role of the interaction between the photoelectron and the source of the field. In the lowest order of the f.s.i. the amplitude is described by Eq.(\ref{9a}), where the interaction $V$ is given by Eqs.(\ref{15}) and (\ref{17}). The integral on the right-hand side of Eq.(\ref{9a}) is determined by small
$f \sim \mu \ll p$, and hence $|{\bf p}-{\bf f}|\gg \mu$. Thus we can employ Eqs.($\ref{16a}$) and $(\ref{19})$. To obtain the accuracy $1/p^2$ it is sufficient to include only the first terms on their right-hand sides. The further procedure just repeats that carried out for the Coulomb field. The analysis
presented in the last paragraph makes the result true for any field with the Coulomb asymptotics.

Finally we find for photoionization of any $ns$ bound state in the field with the Coulomb short range asymptotics \cite{4}
\begin{equation}
\sigma=\frac{16\sqrt{2}\pi^2\alpha(\alpha Z)^2}{3m^{3/2}\omega^{7/2}}n_e e^{-\pi\xi}\psi^2_{ns}(r=0); \quad \xi=m\alpha Z/{\sqrt{2m\omega}}.
\label{22}
\end{equation}
All dependence on the screening is contained in the factor $\psi^2_{ns}(r=0)$.

\section{Asymptotic behavior of the cross sections and analytical properties of the potential}

In previous Section we considered the Coulomb and the screened Coulomb potentials. Both of them have a singularity at $r=0$.
Now we analyze the rectangular well for which the potential $V(r)$ has a discontinuity on the real axis at certain point $r=R$.
The potential can be presented as
\begin{equation}
V(r)=-V_0\theta(r)\theta(R-r); \quad V_0>0.
\label{37}
\end{equation}
Recall that $\theta(x)=1$ for $x>0$ while $\theta(x)=0$ for $x<0$.
Direct calculation provides
$$ V(p)=\frac{4\pi V_0}{p^3}(pR\cos{pR}-\sin{pR}).$$
One can omit the second term in parenthesis since in the asymptotics $pR \gg 1$. We obtain
\begin{equation}
V(p)=\frac{4\pi V_0R}{p^2}\cos{pR}.
\label{38}
\end{equation}
This potential drops as $1/p^2$ at large $p$. Hence the integral (\ref{3d}) for $J(p)$ is dominated by $f \sim \mu$. We need $V({\bf v})$ with
$${\bf v}={\bf p}-{\bf f},$$
to be able to employ Eq. (\ref{3d}). Note that in the well
$\mu \sim 1/R$. Thus $\cos{(vR)}$ in expression for $V({\bf v})=4\pi V_0R\cos{(vR)}/v^2=4\pi V_0R\cos{(vR)}/p^2$ can not be expanded in powers of ${\bf f}$.

Note however that it is sufficient to put
$$ v=p-ft; \quad t={\bf p}{\bf f}/pf, $$
in order to obtain the asymptotics. Thus we obtain
$$J(p)=\frac{4\pi V_0R}{p^2}X(p,R); \quad X(p,R)=\int\frac{d^3f}{(2\pi)^3}\Big(\cos{(pR)}\cos{(ftR)}+\sin{(pR)}\sin{(ftR)}\Big)\psi(f).$$
The second term in the parenthesis vanishes since the function $\psi(f)$ does not depend on direction of ${\bf f}$ for $s$ states.
After integration over $t$ we obtain
$$ X(p,R)=\cos{(pR)}\psi(R),$$
and
\begin{equation}
J(p)=\frac{4\pi V_0R}{p^2}\cos{(pR)}\psi(R); \quad \psi(p)=-\frac{4\pi V_0R}{\omega p^2}\cos{(pR)}\psi(R)
\label{39}
\end{equation}
The photoionization cross section is
\begin{equation}
\sigma=\frac{2^6\pi^2\alpha}{3}\frac{V_0^2R^2}{\omega^2 p^3}\cos^2{(pR)}n_e\psi^2(R); \quad p^2=2m\omega.
\label{39}
\end{equation}
It drops as $\omega^{-7/2}$.

Consider now a simple potential with singularities in the complex plane
\begin{equation}
V(r)=-\frac{U_0}{\pi}\frac{a}{r^2+a^2}; \quad a>0.
\label{43}
\end{equation}
Here $U_0>0$ is a dimensionless constant. One can see that $V(r)\rightarrow -U_0\delta(r)$ for $a\rightarrow 0$, i.e. (\ref{43}) provides a possibility to introduce the zero range potential \cite{7}.

We find for the Foureir transform
\begin{equation}
V(p)=\frac{-4U_0a}{p}\int _{0}^{\infty}dr\frac{r\sin{pr}}{r^2+a^2}=-\frac{2\pi U_0a}{p}\exp{(-pa)}.
\label{46}
\end{equation}
This expression can be obtained by carrying out the integration in complex plane. The potential $V(p)$ is determined by the poles of the integrand in the points $r=\pm ia$.

Recall that we must calculate $\psi(p)=-J(p)/\omega$--see(7). The integral for
\begin{equation}
J(p)=-2\pi U_0a\int\frac{d^3f}{(2\pi)^3}\frac{e^{-v({\bf f})a}}{v(\bf f)}\psi(f); \quad v({\bf f}) =|{\bf p}-{\bf f}|= (p^2-2{\bf p}{\bf f}+f^2)^{1/2},
\label {46a}
\end{equation}
(see (\ref{3d})) is dominated by small $f \ll p$. One can put $v=p$ in the denominator of the integrand. However this can not be done in the power of the exponential factor. To obtain the $p$ dependence of the right hand side we present Eq.($\ref{46a}$)as
$$J(p)=-2\pi\frac{U_0a}{p}I(p); \quad I(p)=\int\frac{d^3f}{(2\pi)^3}e^{-v({\bf f})a}\psi(f).$$
The derivative with respect to $p$ is
$$I'(p)=-a\int\frac{d^3f}{(2\pi)^3}e^{-v({\bf f})a}v'_p({\bf f})\psi(f),$$
with $v'_p=\partial v/\partial p$. One can put $v'_p=1$ with the accuracy $1/p^2$. Thus we can write a simple differential equation
\begin{equation}
I'(p)=-aI(p).
\label{46b}
\end{equation}
with the solution $I(p)=\kappa e^{-pa}$ where $\kappa$ is an unknown constant factor.
Thus
\begin{equation}
J(p)=-2\pi U_0a\kappa\frac{e^{-pa}}{p}.
\label{47}
\end{equation}
Its meaning is prompted by Eq.(21). The $p$ dependence of $J$ is determined by the Fourier transform of the potential $V(r)$. The important distances are $r \sim a $. The wave function at these distances contributes. Hence we can put $\kappa=\psi(r \approx a)$.

Thus we obtained the wave function
$$\psi(p)=2\pi U_0a\kappa\frac{e^{-pa}}{\omega p},$$
� and the cross section
\begin{equation}
\sigma=\frac{16\pi^2\alpha}{3}U_0^2a^2\frac{\exp{(-2pa)}}{\omega^2p}n_e\kappa^2.
\label{47a}
\end{equation}
One can put for estimations $\kappa^2=\psi^2(r=0)$.

We can make a more general conclusion. The cross section for photoionization of a system bound by the field with singularities in the complex plane experiences the exponential drop with $p$. The power of the exponential factor is the imaginary part of the singularity closest to the real axis
multiplied by $2p$.

Consider for illustration the case of modified Poschl-Teller potential which is analyzed in details in the book \cite{6a}
\begin{equation}
V(r)=-\frac{V_0}{ch^2(ar)}; \quad V_0>0;\quad a >0.
\label{47NN}
\end{equation}
The function  $V(r)$ has poles at the points $r=\pm i\pi(2n+1)/2a$ where $n$ is a natural number.
The Fourier transform of this potential for $pa \gg 1$ is determined by the pole $r=i\pi/2a$
\begin{equation}
V(r)=-\frac{\pi^3V_0}{a^3}e^{-\pi p/2a}.
\label{48NN}
\end{equation}
Employing Eqs.(7), (8) and (12) we find expression for the cross section
\begin{equation}
\sigma=\frac{4\pi^6\alpha}{3}\frac{V_0^2}{a^6}\frac{p\exp{(-2pa)}}{\omega^2}n_e\kappa^2.
\label{49NN}
\end{equation}
We can put for estimations $\kappa^2=\psi^2(r=0)$.

Similar analysis can be carried out for the Gaussian potential
\begin{equation}
V(r)=-V_0e^{{-r^2/a^2}}; \quad V_0>0; \quad a>0,
\label{48}
\end{equation}
with the essential singularity in the complex plane.
The Fourier transform is
\begin{equation}
V(p)=-\pi^{3/2}V_0a^3e^{-p^2a^2/4}.
\label{48}
\end{equation}
Proceeding in the same way as in previous case we find
$$\psi(p)=\pi^{3/2}V_0a^3\kappa\frac{e^{-p^2a^2/4}}{\omega},$$
and
\begin{equation}
\sigma=\frac{4\pi^3\alpha}{3}V_0^2a^6\frac{e^{-p^2a^2/2}p}{\omega^2}n_e\kappa^2,
\label{48a}
\end{equation}
where we can put for estimation $\kappa^2=\psi^2(r=0)$.
Hence the cross section drops even faster than $e^{-p^2a^2/2}$.

\section{Photoionization of fullerenes}

Recall that fullerene consist of several dozens of carbon atoms. The $1s$ electrons are bound to their nuclei while the other ones are collectivized.

We consider the photoionization of spherical fullerenes in which the nuclei and the electrons are located in the layer between the spheres of radii $R-\Delta/2$ and
$R+\Delta/2$. Characteristic width of the layer $\Delta$ is of the order $1 a.u.$ while the empirical values of the radii are about $6 a.u.$. We consider their ratio as a small parameter and carry out calculations in the lowest order of $\Delta/R \ll 1$. The model potentials are often employed for description of the fullerene field $V(r)$ which is located mostly inside the thin layer
$ R-\Delta/2 \leq r \leq R+\Delta/2$ \cite{3a}, \cite{3b}.

\subsection{Rectangular well and the Dirac Bubble}
Rectangular well with the limiting points $R_{2,1}=R\pm \Delta/2$ is the prototype of the model fullerene potentials. At $r<R_1$ and at  $r>R_2$ the field vanishes.
The potential $V(r)$ is described by the expression
\begin{equation}
V(r)\ =\ -V_0\theta(r-R_1)\Big(1-\theta(r-R_2)\Big); \quad V_0 > 0.
\label{49}
\end{equation}
Employing Eq.(21) we find
\begin{equation}
\psi(p)=-\frac{4\pi V_0R}{\omega p^2}\Lambda; \quad \Lambda=\cos{(pR_2)}\psi(R_2)-\cos{(pR_1)}\psi(R_1).
\label{39}
\end{equation}
One can express $\Lambda$ in terms of $R$ and $\Delta$.
$$ \Lambda=-\sin{(pR)}\sin{(p\Delta/2)}[\psi(R+\Delta/2)+\psi(R-\Delta/2)]$$
$$+\cos{(pR)}\cos{(p\Delta/2)}[\psi(R+\Delta/2)-\psi(R-\Delta/2].$$
Note that the function $\psi(r)$ changes noticeably in the interval $\Delta$, and thus the difference $\psi(R_2)-\psi(R_1)$ is not small.

The photoionization cross section is
\begin{equation}
\sigma=\frac{2^6\pi^2\alpha}{3}\frac{V_0^2R^2}{\omega^2 p^3}n_e\Lambda^2.
\label{14}
\end{equation}
It drops as $\omega^{-7/2}$.

In the Dirac bubble model
\begin{equation}
V(r)= -U_0\delta(R-r);\quad U_0>0,
\label{49}
\end{equation}
where $U_0$ is a dimensionless constant, the width of the layer $\Delta \rightarrow 0$. Employing Eqs.(21) and (\ref{3}) we obtain immediately
\begin{equation}
\psi(p)=\frac{4\pi U_0R}{\omega p}\sin{(pR)}\psi(R).
\label{50}
\end{equation}
and the cross section
\begin{equation}
\sigma=\frac{2^6}{3}\frac{\pi^2 \alpha U_0^2R^2}{\omega^2p}\sin^2{(pR)}n_e\psi^2(R); \quad p^2=2m\omega.
\label{50}
\end{equation}
The latter drops as  $\omega^{-5/2}$.
The expression for the cross section was found earlier \cite{8} by solving the wave equation.

The photoionization cross section for the Dirac bubble drops slower than that for the rectangular well. The difference is due to the different character of the potentials at the points of discontinuity. In the former case $V(r)$ is infinite at $r=R$. It experiences the finite jumps at $r=R_{1,2}$ in the latter case. This causes different character  of singularities of the wave function $\psi(r)$ at the discontinuity points. In the case of the Dirac bubble potential the derivative $\psi'(r)$ experiences a jump at $r=R$ \cite {7}. It is a continuous  function at the points $r=R_{1,2}$ for the rectangular well\cite{6} while the second derivative experiences a jump. To demonstrate that the wave function for the Dirac bubble undergoes a jump
at $r=R$ consider the wave equation for the function $\chi(r)=r\psi(r)$
\begin{equation}
 -\frac{\chi''(r)}{2m}+V(r)\chi(r)=\varepsilon_B\chi(r); \quad V(r)= -U_0\delta(r-R).
\label{51}
\end{equation}
� $V(r)= -U_0\delta(r-R)$. Integrating both sides over a small interval near the point $r=R$, i.e. $R-\delta \leq r \leq R+\delta$, we obtain
$$\chi'(R+\delta)-\chi'(R-\delta)=-2mU_0\chi(R).$$
This is the explicit expression for the jump.

The  different character  of singularities of the wave function $\psi(r)$ at the discontinuity points leads to different $p$ dependence of the wave functions. Presenting
$$\psi(p)=\int d^3r \psi(r)e^{-i{\bf p}{\bf r}}=\frac{4\pi}{p}\int_0^{\infty}dr \sin{(pr)}\chi(r).$$
we find for the Dirac bubble potential
$$
\psi(p)=\frac{4\pi}{p}\Big[\int_0^{R_{-}}dr\sin{pr}\chi(r)+\int_{R_{+}}^{\infty}dr\sin{pr}\chi(r)\Big]; \quad R_{\pm}=R\pm \delta; \quad \delta \rightarrow 0. $$
After two integrations by parts we obtain
\begin{equation}
\psi(p)=\frac{4\pi}{p^3}R\sin{(pR)}[\psi'(R_{-})-\psi'(R_{+})]
\label{52}
\end{equation}
$$-\frac{4\pi}{p^3}\Big[\int_0^{R_{-}}dr\sin{pr}\chi^{(2)}(r)+\int_{R_{+}}^{\infty}dr\sin{pr}\chi^{(2)}(r)\Big].$$
The asymptotics at $ p \rightarrow \infty$ is determined by the first term on the right-hand side. In the case of the rectangular well the derivatives $\psi'(r)$ are continuous at the points
$R_{1,2}$. The asymptotics can be obtained by integration by  parts of the second term with putting $R_{\pm}=R_1\pm \delta$ and
$R_{\pm}=R_2\pm\delta$. Hence the wave function $\psi(p)$ for the rectangular well obtains additional factor $1/p$ compared with the Dirac bubble case.
This provides additional factor $1/\omega$ in the photoionization cross section.
Note that although Eq.(54) does not contain explicit dependence on $U_0$, the jump $\psi'(R_{-})-\psi'(R_{+})$ is
proportional to $U_0$.

\subsection{Jellium model}

The jellium model is often employed in nowadays calculations (see, e.g. \cite{9},\cite{10}). In this model the positive charge of the fullerene core, consisting of nuclei and the internal ($1s$) electrons  is assumed to be uniformly distributed inside the layer with the width $\Delta$.  At  $r<R_1$ the field $V(r)$ is constant while at $r>R_2$ it exhibits the Coulomb behavior. We denote the field of the core $V(r)=V_1(r)$ at $0\leq r <R_1$, $V(r)=V_2(r)$ at $R_1\leq r  \leq R_2$  and $V(r)=V_3(r)$ at $ r > R_2$. The potentials $V_{1,2,3}$ are
\begin{equation}
V_1(r)\ =-U_0\frac{3}{2}\frac{R_2^2-R_1^2}{R_2^3-R_1^3} ; \quad V_2(r)=-\frac {U_0}{2(R_2^3-R_1^3)}\Big(3R_2^2-r^2(1+\frac{2R_1^3}{r^3})\Big);
\label{53}
\end{equation}
$$V_3(r)=-\frac{U_0}{r}; \quad U_0>0.$$
Here $U_0$ is the dimensionless constant. The potential $V(r)$ determined by Eq.(\ref{53}) and its derivative $V'(r)$ are continuous at the real axis, i.e. $V_1(R_1)=V_2(R_1)$;
$V_2(R_2)=V_3(R_2)$, �  $V_1'(R_1)=V_2'(R_1)$; $V_2'(R_2)=V_3'(R_2)$. The second derivative  $V^{(2)}(r)$ experiences jumps at $r=R_{1,2}$.
This determines the asymptotics of the cross section.

We combine Eq.(7) and Eq.(21) to find  the asymptotics of the wave function
\begin{equation}
\psi(p)=-\frac{1}{\omega}\int d^3r e^{-i{\bf f}{\bf r}}V(r)\psi(r).
\label{54}
\end{equation}
The angular integration provides
\begin{equation}
\psi(p)=-\frac{4\pi}{\omega p}\int_0^{\infty}dr\sin{(pr)}\varphi(r); \quad \varphi(r)=rV(r)\psi(r)=V(r)\chi(r).
\label{55}
\end{equation}
Here $\int_0^{\infty}dr=\int_0^{R_1-\delta}dr+\int_{R_1+\delta}^{R_2-\delta}dr+\int_{R_2+\delta}^{\infty}dr, \delta \rightarrow 0$.
Singularities of the higher derivatives of the integrand do no influence the results of two first integrations by parts
\begin{equation}
\psi(p)=\frac{4\pi}{\omega p^3}\int_0^{\infty}dr\sin{(pr)}\varphi^{(2)}(r).
\label{56}
\end{equation}
Next integration by parts provides
\begin{equation}
\psi(p)=-\frac{4\pi}{\omega p^4}\sum_{i=1,2}\lambda_i+\frac{4\pi}{\omega p^4}\int_0^{\infty}dr\cos{(pr)}\varphi^{(3)}(r).
\label{57}
\end{equation}
Here
$$\lambda_i=\chi(R_i)\cos{(pR_i)}[V^{(2)}(R_i-\delta)-V^{(2)}(R_i+\delta)],$$
are the jumps of the integrand at the points  $R_{1,2}$. One can see from Eqs. (\ref{51}) that the derivatives $\psi^{(2)}$
do not experience jumps at the points $R_{1,2}$. Asymptotics of the wave function is determined by the first term on the right hand side of Eq.(\ref{57}). The second term provides the contribution quenched as $1/p$ in further integrations by parts.

Asymptotics of the photoionization cross section is
\begin{equation}
\sigma=\frac{64\pi^2\alpha}{3\omega^2p^7}n_e\Lambda^2 ; \quad \Lambda=\lambda_1+\lambda_2.
\label{58}
\end{equation}
It drops as  $\omega^{-11/2}$.

\subsection{The Lorentz bubble }

This potential is determined by the expression
\begin{equation}
V(r) = -\frac{U_0}{\pi}\frac{a}{(r-R)^2+a^2}.
\label{59}
\end{equation}
At $R=0$ this is just the potential described by Eq.(35). The Dirac bubble can be viewed as the limiting case of the Lorentz bubble at
$a \rightarrow 0$. It is reasonable to consider this potential with $a \sim \Delta \ll R$ in fullerene physics..

The Fourier transform for the Lorentz bubble can be written as
\begin{equation}
V(p)=V_A+V_B; \quad V_{A,B}=-4\frac{a}{p}X_{A,B},
\label{60}
\end{equation}
with
\begin{equation}
X_A(p)=U_0(-\frac{\partial}{\partial p})\cos{(pR)}\int_{-\infty}^{\infty}dx\cos{(px)}u(x);
\label{18}
\end{equation}
$$X_B(p)=U_0\frac{\partial}{\partial p}\int_{R}^{\infty}dx\cos{((p(x-R))}u(x) \quad u(x)=\frac{1}{x^2+a^2}.$$
Here we changed the variable of integration introducing $x=r-R$.
The wave function can be presented as
\begin{equation}
\psi(p)=\psi_A(p)+\psi_B(p)
\label{63}
\end{equation}
with the two terms on the right-hand side corresponding to those on the right-hand side of Eq.(\ref{60}).

In the lowest order of expansion in powers of $a/R \ll 1$ we obtain the potential
\begin{equation}
V_A(p)=-4\pi \frac{U_0R}{p}e^{-pa}\sin{pR}.
\label{64}
\end{equation}
The leading contribution to $V_A(p)$, comes from $x \sim a$.
Here $|r-R| \sim a$ and the electron density reaches its largest values.

Introducing $r'=x-R$ we present also
$$V_B(p)=\frac{4U_0a}{p}\int_{0}^{\infty}dr'\frac{r'\sin{pr'}}{(R+r')^{2}+a^2}.$$
The integral on the right hand side is determined by small $ r' \sim 1/p$. The asymptotical expansion of the integral in powers of $(pR)^{-1}$ provides
$$V_B(p)=\frac{16U_0aR}{p^3(R^2+a^2)^2}.$$
In the lowest order of $a^2/R^2$ expansion we obtain
\begin{equation}
V_B(p)=\frac{16U_0a}{(pR)^3},
\label{65}
\end{equation}

Now we calculate the wave functions $\psi_A(p)$ and $\psi_B(p)$ introduced by Eq.(64). Combination of Eq.(7), Eq.(8) and Eq.(63) gives
\begin{equation}
\psi_A(p)=\frac{4\pi U_0R}{\omega}\int\frac{d^3f}{(2\pi)^3}\frac{e^{-va}}{v}\sin {(vR)}\psi(f); \quad
v=|{\bf p}-{\bf f}|.
\label{66}
\end{equation}
We proceed in the same way as we did in calculations for the rectangular well in Sec.5, putting
$v=p-ft$ for calculation of asymptotics
\begin{equation}
\psi_A(p)=\frac{4\pi U_0R}{\omega}X(p,R),
\label{67}
\end{equation}
$$X(p,R)=\int\frac{d^3f}{(2\pi)^3}\frac {e^{-va}}{v}\Big(\sin {(pR)}\cos(ftR)-\cos{(pR)}\sin{(ftR)}\Big)\psi(f).$$
Due to the expression in the parenthesis $X(p,R)$ is determined by the region  in which  $|ft| \sim 1/R$. This enables to put
$v=p$ and $e^{-va}=e^{-pa}$ � in the first factor of the integrand expressing $X(p,R)$.
Since the wave function $\psi(f)$ does not depend on the directions of ${\bf f}$ the angular integration provides
$$X(p,R)=\frac{e^{-pa}\sin{(pR)}}{p}\int\frac{d^3f}{(2\pi)^3}\frac {\sin{(fR)}}{fR}\psi(f).$$
This is just
\begin{equation}
X(p,R)=\frac{e^{-pa}\sin{(pR)}}{p}\psi(R)
\label{68}
\end{equation}
Thus
\begin{equation}
\psi_A(p)=4\pi U_0R\frac{e^{-pa}\sin {(pR)}}{\omega p}\psi(R).
\label{69}
\end{equation}

Employing Eq.(9) we obtain
\begin{equation}
\psi_B(p)=-\frac{16U_0a}{\omega p(pR)^3}\psi(r=0),
\label{70}
\end{equation}
From the formal point of view, $\psi_B(p)$ determines the asymptotics of the wave function. Thus the asymptotic cross section
\begin{equation}
\sigma_B=\frac{2^{10}\alpha}{3}\frac{U_0^2a^2}{R^6}\frac{\psi^2(r=0)}{\omega^2p^7},
\label{71}
\end{equation}
drops $\omega^{-11/2}$.

Note that the cross section $\sigma_B(p)$ is proportional to $\psi^2(r=0)$ while we have seen the potential $V_B(p)$ to be determined by small distances $r' \sim 1/p $. They correspond to the distances  $r$ from the center of the fullerene close to $2R$. These two statements do not contradict each other. The potential determined by Eq.$(\ref{59})$,is invariant under the transformation  $r'=2R-r$. Thus Eq.$(\ref{51})$ provides
$\chi(2R-r)=C\chi(r)$ where $C$ is constant. In particular, at  $r \ll R$ we can put  $\chi(2R)=C\chi(r)$. At any $x$ and $\delta \ll x$ the value $n_x=\int_x^{x+\delta}dy \chi^2(y)$ determines the part of the electron density located in the spherical layer with the width $x$. For $r \ll R$
we can write $\chi^2(2R)=C^2\chi^2(r)$. Thus the part of the electron density near the spherical surface with the radius $2R$ is the same as near the surface with the small radius $r$.

 We shall see that the power drop of the cross section is reached at rather large values of the photoelectron energies. Here the cross section becomes unobservably small. It is the region where the wave function  $\psi(p)$ is determined by the contribution
$\psi_A(p)$, which is of practical interest. The observable cross section is
\begin{equation}
\sigma=\frac{64}{3}\frac{\alpha\pi^2U_0^2R^2}{\omega^2p}e^{-2pa}\sin^2{(pR)}n_e\psi^2(R).
\label{72}
\end{equation}
It differs from that for the Dirac bubble potential given by Eq. (52) by the exponential factor $e^{-2pa}$.

The contribution $\sigma_A$ dominates since  $|\psi_B(p)| > |\psi_A(p)|$ only if
\begin{equation}
\frac{e^{pa}}{(pa)^3}>\frac{3}{4}\frac{R^4}{a^4}\tau; \quad \tau=\frac{|\psi(R)|}{|\psi(0)|}.
\label{73}
\end{equation}
The ratio $\tau \gg 1$ (recall that the electron density is located mainly at $|r-R| \ll R$) can be obtained in the simplest models. We employ the result for the Dirac bubble potential $\tau=e^{\mu R}/{2\mu R}$ ($\mu=(2mI_B)^{1/2}$)\cite{8}, \cite {4}. At characteristic $R=6 ~a.u.$, $I_B=1~ Ry$ we find $\tau \approx 34$. For $a=1 a.u.$ the relation
(74) becomes true at $pa >19$, corresponding to photoelectron energies  $\varepsilon >5$ keV. The cross section determined by Eq. (73) at $pa=19$ is quenched more than $10^8$ times compared to that at $pa=3$ ($\varepsilon \approx 100$ eV). It has no chances to be observed.

\subsection {Gaussian type potential}

A potential with  sharp  maximum at $r=R$ was suggested in \cite{11}
\begin{equation}
V(r) = -\frac{V_0}{\pi}\exp{\Big(\frac{-(r-R)^2}{a^2}\Big)}; \quad V_0>0.
\label{75}
\end{equation}
with $a \sim \Delta \ll R$.

We proceed in the same way as we did for the Lorentz bubble. Similar to Eq.(62) we present the Fourier transform of the potential as $V(p)=V_A(p)+V_B(p)$, where $V_{A,B}(p)=-4V_0X_{A,B}(p)/p$. Here $X_{A,B}$ are expressed by Eqs.(63) with $u(x)=e^{-x^2/a^2}$ and $U_0=1$.
We find the contribution determined by $|r-R| \sim a$
\begin{equation}
V_A(p)=\sqrt{\pi}V_0Ra\frac{e^{-p^2a^2/4}}{p}\Big(\sin{pR}+\frac{a}{2R}pa\cos{pR}\Big).
\label{77}
\end{equation}
While $pa \ll 2R/a$, i.e. $\varepsilon \ll 2$ keV for the characteristic values of the parameters assumed in 6.3, the right-hand side of Eq.(76) is dominated by the first term. We find for the observable photoionization cross section
\begin{equation}
\sigma=\frac{64\alpha \pi}{3}\frac{V_0^2R^2a^2}{\omega^2p}e^{-p^2a^2/2}\sin^2{(pR)}n_e\psi^2(R),
\label{78}
\end{equation}

The energy interval determined by the condition $pa \ll 2R/a$ is the only one, where the cross section is of practical interest.
One can see from Eq.(77) that the cross section becomes $4\cdot 10^4$ times smaller while the photoelectron energies change from 100 eV to 300 eV. It becomes many orders of magnitude smaller when $pa \sim 2R/a$.

The leading term of the asymptotic expansion for the potential
\begin{equation}
V_B(p)=\frac{16V_0R}{p^3a^2}e^{-R^2/a^2}.
\label{79}
\end{equation}
determines the cross section
\begin{equation}
\sigma_B=\frac{16}{3}\frac{\alpha V_0^2R^2}{a^4\omega^8p^5}e^{-2R^2/a^2}n_e\psi^2(r=0).
\label{80}
\end{equation}
From the formal point of view this is the asymptotic expression for the cross section. However it describes the total photoionization cross section only if
$$ \frac{e^y}{y^2} > 2\sqrt{\pi}\frac{|\psi(R)|}{|\psi(0)|}\frac{a}{R}\exp ({R^2}/{a^2}); \quad y=\frac {p^2a^2}{4},$$
 i.e. at $ y>46$ ; $\varepsilon > 2.5 $ keV. In this energy region the cross section is too small to be observed. Eq.(79) may be of interest for the systems with smaller values of the ratio $R/a$ than for the fullerenes known today.

\section{Summary}

 In the present paper we considered the systems bound by central potential $V(r)$. We demonstrated that the energy behavior of the photoionization cross section at large but nonrelativistic photoelectron energies can be obtained without solving the wave equation. Asymptotics of the cross section is expressed in terms of the Fourier transform of the potential $V(p)$ as
$\sigma(\omega)=4\alpha p|V(p)|^2\kappa^2/(3\omega^2)$ (see Eq.(12)), where the parameter  $\kappa$ does not depend on the photon energy.
In the simplest case $\kappa^2=|\psi(r=0)|^2$.

We showed that our expressions are true for the well studied case of the Coulomb field. We found the asymptotics of the photoionization cross section for the screened Coulomb field. We demonstrated that the leading corrections to the asymptotic behavior are described by the same Stobbe factor as in the case of the Coulomb field.

The energy behavior of photoionization cross section is found to be determined by the analytical properties of the potential. The potentials with singularities on the real axis lead to the power drop of the cross sections. The potentials with singularities in the complex plane correspond to exponential drop of the cross sections. The potential with essential singularity in the complex plane provide still faster drop of the cross section.

The model potentials are widely used in studies of the fullerene physics. The results of the paper can be applied for analysis of the high energy photoionization of fullerenes. We compared the energy dependence of photoionization cross sections in three models with singularities on the real axis. The slowest drop of the cross section $\omega^{-5/2}$ takes place in the Dirac bubble model with the infinite jump of the potential. The cross section drops as  $\omega^{-7/2}$ in the rectangular well field, where the potential $V(r)$ experiences finite jumps. The photoionization cross section behaves as $\omega^{-11/2}$ in the jellium model. The faster drop in this case is caused by a somewhat more smooth behavior of the potential. Only its second derivatives suffer jumps.

If the fullerenes are described by the model potentials with singularities in the complex plane such as the Lorenz bubble or the Gaussian-type potential, the asymptotics, from the formal point of view, experience a power drop $\omega^{-11/2}$ being determined  by the small distances  $ r \sim 1/p$. However such behavior is reached at the photoelectron energies which are so large that the cross section becomes unobservably small. The preasymptotic behavior is of real interest. It is determined by the distance region where the electron density reaches its largest values. Here the energy dependence of the photoionization cross section is described by the dropping exponent in the case of the Lorentz potential and by the
Gaussian curve in the case of the Gaussian--type potential.

\end{document}